\documentstyle[preprint,prl,aps]{revtex}
\begin{document}
\draft
\title{Collective Coulomb Blockade in an Array of Quantum Dots:\\
A Mott-Hubbard Approach}
\author{C.~A.~Stafford and S.~Das Sarma}
\address{Center for Superconductivity Research, Department of Physics, \\
University of Maryland, College Park, Maryland 20742}
\date{Received 25 August 1993}
\maketitle
\begin{abstract}
We investigate
the electron addition spectrum in a class of Hubbard-like models which
describe arrays of coupled quantum dots.
Interdot tunneling leads to a sequence
of two phase transitions separating a region of collective Coulomb
blockade from a region where the Coulomb blockade of individual dots is
maintained and a region where the Coulomb blockade is destroyed
altogether.  Observable experimental consequences of our theory are discussed.
\end{abstract}

%\tighten
\narrowtext
Arrays of coupled quantum dots \cite{dotarrays,1dcrystal}
provide a novel system in which one
can study ``solid-state'' physics on a much lower energy scale.
Because of the dominant role played by electron-electron interactions
in quantum dots, as manifested in the phenomenon of Coulomb
blockade \cite{blockade},
it is expected that quantum dot arrays should
mimic the physics of strongly-correlated lattice systems, and should
therefore provide rather good
experimental testing grounds for studying the predictions of
Hubbard-type strongly-correlated electron interaction models.
To date, theoretical treatments of quantum dot arrays
have focused largely on arrays of metallic dots
\cite{classicaldots}, where the discreteness of the dot
energy levels can be neglected, or arrays with nearly transparent
barriers between dots \cite{noninteractingdots},
where Coulomb blockade is absent.
In this Letter, we investigate the electron addition spectrum in
arrays of coupled semiconductor quantum dots in the
Coulomb blockade regime and its vicinity.
Our results directly reflect the intricacies of Mott-Hubbard
correlations, and, in particular, show that various interaction-driven
phase transitions \cite{mott,meanfield,andyandi,crit}
should be observable in suitably fabricated
quantum dot arrays.

Motivated by the success in describing Coulomb blockade in a single quantum
dot using an Anderson-type Hamiltonian \cite{onedot}, we
employ a generalized Hubbard model to account for the effects of quantum
confinement, intradot Coulomb interactions,
and interdot tunneling in a quantum dot array.  We find a remarkably
rich phase diagram for quantum dot arrays which generically exhibits
three phases:  for weak interdot tunneling, the quantum dot states are
split into minibands but the Coulomb blockade of individual dots is
maintained; for intermediate tunneling, the Coulomb blockade of
individual dots is destroyed but there remains an energy gap due to
interdot correlations (collective Coulomb blockade);
and for strong tunneling, the Coulomb blockade is destroyed
altogether.  Our results provide new insight into an early experiment on
transport through a one-dimensional (1D)
array of quantum dots \cite{1dcrystal}, as well as
a possible explanation for anamolous
degeneracies observed in the addition spectrum of a disordered quantum dot
with several local minima \cite{ray}.

The system we wish to model consists of a linear or square array of
identical quantum dots of the type described in Ref.~\cite{ray2},
separated from a metallic backgate by a thin insulating layer.
We represent the Coulomb interaction between electrons on the same dot
by an energy $U=e^2/C$, the capacitive charging energy of the dot
\cite{blockade},
and assume that interdot Coulomb interactions are screened by the backgate.
The single-particle energy levels in the confining potential of an
isolated dot are denoted by $\varepsilon_{\alpha}$, $\alpha=1,\ldots,M$
($\alpha$ specifies the quantum state of both the orbital and spin
degrees of freedom),
which we take to be nondegenerate with level spacing $\Delta$.
%(we set $\varepsilon_1=0$).
The dominant effect of interdot coupling is to introduce a tunneling matrix
element $t_{\alpha}$ between equivalent single-particle states in
nearest-neighbor dots.  We neglect the tunneling matrix elements between
nonequivalent states; this is the usual tight-binding approximation, and
is justified for nearly identical dots provided $t_{\alpha}$ is not too
large.  The Hamiltonian is
\begin{equation}
\hat{H} = -\sum_{\stackrel{\scriptstyle
\langle {\bf i},{\bf j} \rangle}{\alpha}}
\left(t_{\alpha} \, \hat{c}_{{\bf i} \alpha}^{\dagger} \hat{c}_{{\bf j}
\alpha} + \mbox{H.c.}\right) + \sum_{{\bf j},\alpha} \varepsilon_{\alpha} \,
\hat{c}_{{\bf j} \alpha}^{\dagger} \hat{c}_{{\bf j} \alpha} +
\frac{U}{2} \sum_{\bf j} \hat{n}_{\bf j} (\hat{n}_{\bf j} -1),
\label{hubham}
\end{equation}
%\FL
%\begin{eqnarray}
%\hat{H} & = & -\sum_{\stackrel{\scriptstyle
%\langle {\bf i},{\bf j} \rangle}{\alpha}}
%\left(t_{\alpha} \, \hat{c}_{{\bf i} \alpha}^{\dagger} \hat{c}_{{\bf j}
%\alpha} + \mbox{H.c.}\right) + \sum_{{\bf j},\alpha}
%\varepsilon_{\alpha} \,
%\hat{c}_{{\bf j} \alpha}^{\dagger} \hat{c}_{{\bf j} \alpha} \nonumber\\
%\mbox{} & \mbox{} & \;\;\;\;\;\;\;\;\;\;\;\;\;\;\;\;\;\;\;\;\;\;\;\;
%\mbox{ } + \frac{U}{2} \sum_{\bf j} \hat{n}_{\bf j} (\hat{n}_{\bf j} -1),
%\label{hubham}
%\end{eqnarray}
where ${\bf i}, {\bf j}$ are vectors of integers labeling the positions
of the dots,
$\hat{c}^{\dagger}_{{\bf j} \alpha}$ is the creation operator for an
electron in state $\alpha$ of the $\bf j$th dot,
$\hat{n}_{\bf j} \equiv \sum_{\alpha} \hat{c}_{{\bf j} \alpha}^{\dagger}
\hat{c}_{{\bf j} \alpha}$, and the sum over $\langle {\bf i},{\bf j}
\rangle$ is over nearest neighbors only.
In the following, we consider arrays of 4 and 6 quantum dots with $M=2$,
3, and 4 single-particle energy levels per dot.

In this Letter, we calculate the equilibrium electron addition spectrum of
(\ref{hubham}), $\partial \langle \hat{N} \rangle /\partial \mu =
kT\, \partial^2 \ln {\cal Z}/\partial \mu^2$,
by diagonalizing $\hat{H}$ and evaluating the grand partition function
${\cal Z} = \mbox{Tr}\{\exp[-(\hat{H}-\mu \hat{N})/kT]\}$,
where $\hat{N}=\sum_{\bf j} \hat{n}_{\bf j}$, $\mu$ is the chemical
potential, and $\langle \mbox{ } \rangle$ denotes the thermal average.
For the largest array considered, $\cal Z$ involved a sum over 65,536 states.
Assuming that the tunneling rate from the backgate
is much less than the temperature $kT$,
the differential self-capacitance of the experimental quantum dot system
is given by
$\partial Q/\partial V = e^2 \partial \langle \hat{N} \rangle /\partial \mu$,
which is directly measured by the technique of single-electron capacitance
spectroscopy \cite{ray,ray2}.

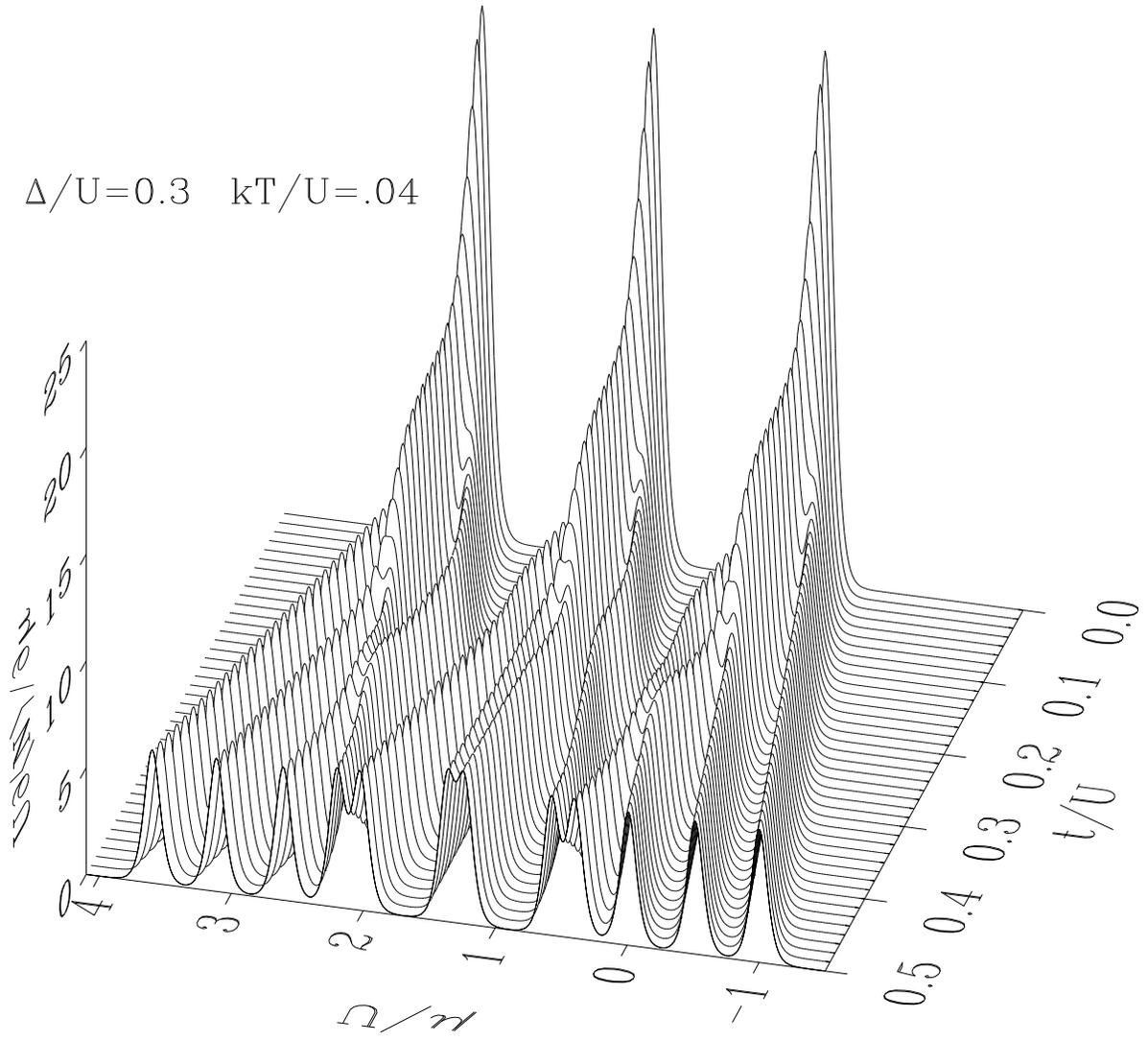
\begin{figure}
\caption{A plot showing the evolution of the electron addition spectrum
$\partial \langle \hat{N} \rangle / \partial \mu$
as a function of the interdot tunneling matrix element $t$
for an array of four quantum dots with three single-particle energy
levels per dot, arranged in a square.}
\label{fig1}
\end{figure}

We first consider an array of four quantum dots arranged in a square
with three single-particle energy levels per dot, and study the
evolution of the addition spectrum as the interdot tunneling is
increased (see Fig.~\ref{fig1}).  Here we have set the tunneling matrix
elements for all three quantum levels equal to $t$ and have taken
$\Delta=0.3 U$ and $kT=.04 U$.  At $t=0$, the behavior characteristic of
isolated dots \cite{blockade} is evident in Fig.~\ref{fig1}; namely,
three peaks separated by the Coulomb blockade energy $U+\Delta$.  Each
peak at $t=0$ represents the addition of four electrons to the array
(one to each quantum dot).  As $t$ is increased, the quantum dot states
are split into minibands and each capacitance peak is split into three
peaks, the central peak representing the addition of two electrons at
the same value of the chemical potential.  When $t \sim \Delta$, this
degeneracy is lifted and one can see three Hubbard minibands, each
composed of four states.  Finally, when $t \sim U/2$ the energy gap
between the minibands is no longer discernible.

Figure \ref{fig1} shows evidence of three distinct phases in the quantum
dot array, which we will now analyze in detail.  In the weak-tunneling
phase, characterized by the degeneracies in the addition spectrum,
$\partial \langle \hat{N} \rangle / \partial \mu$ has peaks at zero
temperature at $\mu = -2t$, $0(\times 2)$, and $2t$, with this pattern
repeated, centered at $U + \Delta$ and at $2(U + \Delta)$.  The energies
of the lowest miniband are those of noninteracting electrons in a
tight-binding band with four lattice sites, $E=-2t\cos k$, $k=0,\pm
\pi/2,\pi$.  This reflects the fact that at sufficiently small $t$
there is no admixture of the higher single-particle states $\varepsilon_2$
and $\varepsilon_3$ in the many-body ground state,
so that the Pauli principle prevents two or more
electrons from occupying the same dot, thus negating the interaction
term in Eq.~(\ref{hubham}).  The minibands centered at $U + \Delta$ and
$2(U + \Delta)$ are identical, but occur when the states $\varepsilon_1$
and $\varepsilon_2$ are completely filled.  This weak-tunneling phase is
analogous to the ferromagnetic phase of the Hubbard model in a strong
magnetic field, the level splitting $\Delta$ playing the role of the field.

As $t$ is increased, it becomes energetically favorable to admix
the higher single-particle states $\varepsilon_2$ and $\varepsilon_3$ in
the many-body ground state---thereby allowing multiple occupancy of the
quantum dots---in order to lower the kinetic energy of the system.
The degeneracies present in the weak-tunneling phase are then lifted by
the interaction term in Eq.~(\ref{hubham}).
In Fig.~\ref{fig1}, this transition occurs when $t \sim
\Delta$.  For the case of an infinite 1D array of quantum dots with
$M=2$ levels per dot, this phase transition is equivalent to the
ferromagnetic/antiferromagnetic phase transition of the 1D Hubbard model
in a magnetic field \cite{crit}, and in the
limit $U \gg \Delta$ the critical value of the tunneling matrix element
is given by \cite{crit}
\begin{equation}
t_{c}^2 = \frac{\pi \Delta U}{4 (2\pi n - \sin 2 \pi n)},
\label{tc}
\end{equation}
where $n < 1$ is the mean number of electrons per dot.  We find that this
transition is qualitatively similar for $M=3$ and 4, and in finite arrays
(although it is of course not a true phase transition in a finite array).
This transition represents a breakdown of
the Coulomb blockade of individual quantum dots:  when $t < t_c$, there
is no multiple occupancy of the quantum dots in the lowest miniband,
while for $t > t_c$, there is a nonzero amplitude for multiply
occupied dots in the lowest miniband.  (It is
crucial that the $\varepsilon_{\alpha}$ not be spin-degenerate,
otherwise $t_c =0$.)
Prediction of this observable ``phase transition'' in quantum dot arrays
is one of our important new results.

Despite the destruction of
Coulomb blockade in the individual quantum dots when $t > t_c$,
there is still an
energy gap between the minibands above this transition.  This energy gap
is a collective effect which we refer to as ``collective Coulomb
blockade'' (CCB), and is analogous to the energy gap in a Mott insulator
\cite{mott,meanfield}.  The CCB regime is characterized by strong interdot
correlations which are analogous to the antiferromagnetic correlations
in Mott insulators \cite{pwa}; {\it i.e.}, occupancy of the state
$\varepsilon_{\alpha}$ in one quantum dot is anticorrelated with the
occupancy of that state in its nearest neighbors.

When $t$ is increased still further, the energy gap between minibands in
Fig.~\ref{fig1} collapses.  The breakdown of CCB \cite{spindegen}
in the strong-tunneling
regime is directly analogous to the Mott-Hubbard insulator-metal transition
\cite{mott,meanfield,andyandi}.
For $M=2$ single-particle energy levels per dot,
this transition occurs at $t/U=\infty$ for a 1D array \cite{lwu}, but
for $M>2$, the transition is expected to occur at a finite value of $t$
due to the absence of Fermi-surface nesting \cite{meanfield,sun};
Fig.~\ref{fig1} is consistent with the critical value
obtained in Ref.~\cite{sun} for an infinite array with $M=3$, $t_c/U=0.39$
\{although the metal-insulator transition studied in Ref.~\cite{sun} occurs
in an unspecified model which is merely analogous to Eq.~(\ref{hubham})\}.
The signature of this ``metal-insulator'' transition in the electron
addition spectrum is relatively insensitive to finite-size effects
(which can have a drastic effect on the conductivity of the system
\cite{andyandi}).

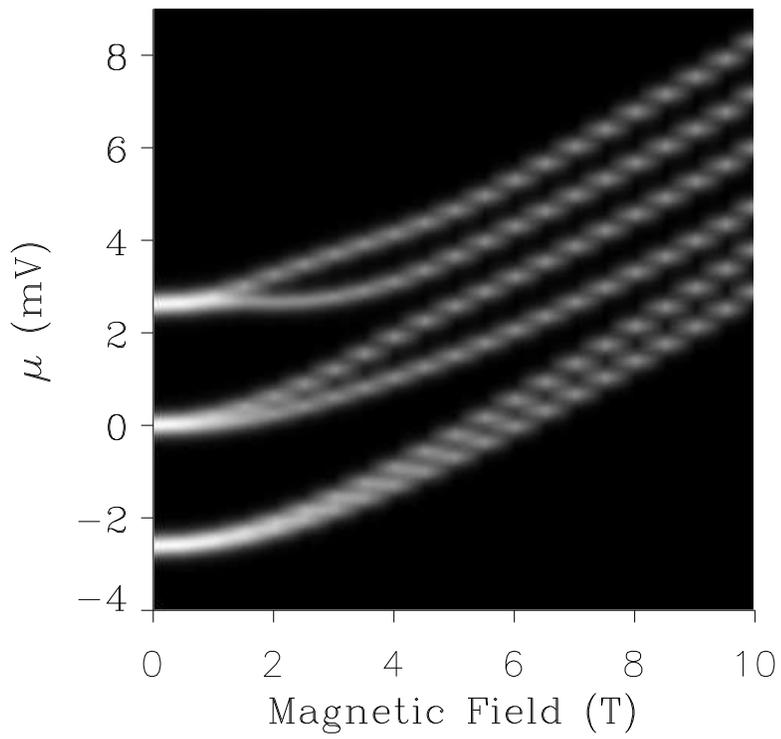
\begin{figure}
\caption{A gray scale image of the electron addition spectrum
%$\partial \langle \hat{N} \rangle / \partial \mu$
versus magnetic field in the lowest miniband of a ring
of six quantum dots with two single-particle energy levels per
dot.  Here $U = 10\mbox{mV}$, $\Delta=3\mbox{mV}$, $t=1.5\mbox{mV}$,
and $T=1.1\mbox{K}$.  The radius of the ring is 62nm
and the effective mass for bulk GaAs has been used.}
\label{fig2}
\end{figure}

Figure \ref{fig2} shows the effects of disorder and magnetic field on
the degeneracies which occur in the addition spectrum
in the weak-tunneling regime.
$\partial \langle \hat{N} \rangle / \partial \mu$
is plotted in gray scale, where white represents the peaks.
The data are for the lowest miniband of a singly-connected array
of six quantum dots with two single-particle energy levels per dot.
The area enclosed by the ``ring'' of dots is $(110\mbox{nm})^2$, and
the system parameters are $U = 10\mbox{mV}$, $\Delta=3\mbox{mV}$,
$t=1.5\mbox{mV}$, and $T=1.1\mbox{K}$.
Disorder with a magnitude $\sim 0.2\mbox{mV}$
is included in the on-site energies,
and one of the $t$'s is taken to have a negative sign in zero field.
The magnetic field modifies the interdot tunneling matrix
elements in Eq.~(\ref{hubham}) by a Peierls phase factor (we
neglect modifications to the magnitude of $t$).  The
modification of the quantum confinement by the magnetic field may be
taken into account \cite{ray2} by setting
$\varepsilon_1= [\Delta^2 + (\omega_c/2)^2]^{1/2} - \Delta$,
$\varepsilon_2= 2[\Delta^2 + (\omega_c/2)^2]^{1/2} - \Delta -\omega_c/2$,
where $\omega_c$ is the cyclotron energy (the effective mass of bulk
GaAs has been used).  The Zeeman term is only .09mV at 10T, and is neglected.
Disorder mixes the states in the miniband with $\pm k$
and splits the degeneracies present at zero field by an amount
$\sim (\overline{\varepsilon_{1}^2} - \overline{\varepsilon_1}^2)^{1/2}$
(disorder in the hopping matrix elements has a similar effect).
This splitting is obscured in Fig.~\ref{fig2} by
the finite electron temperature.  The magnetic field has two quite
different effects: it explicitly
breaks time-reversal invariance, leading to a splitting proportional
to the field; in addition, because the splitting between $\varepsilon_1$ and
$\varepsilon_2$ is a decreasing function of magnetic field,
the magnetic field can actually drive the system
into the CCB regime, leading to an abrupt splitting.  In Fig.~\ref{fig2},
this field-induced phase transition is clearly visible in the highest
doublet at $\sim 1\mbox{T}$, and in the intermediate doublet at $\sim 2
\mbox{T}$, but is absent in the lowest doublet, in agreement with the
density dependence of Eq.~(\ref{tc}).

The experimental capacitance
spectrum obtained by Ashoori {\it et al}.\ \cite{ray} for a disordered
quantum dot with several local minima---which can be thought of as a random
array of small quantum dots between 130\AA~and 360\AA~in size---exhibits
degeneracies much like those in Fig.~\ref{fig2},
which are split by a magnetic field.
If these degeneracies are to be ascribed to the above mechanism,
it would require that the local minima be arranged in
a nearly symmetrical pattern
(it is not necessary that they form a ring---similar degeneracies occur
in other symmetrical arrays.)  This possibility is consistent with the
fact that the degeneracies disappeared after the sample was thermally
cycled \cite{ray}.  Of course,
it is crucial for this interpretation that the energy levels in the
local minima be non-spin-degenerate, which would require spin-orbit
coupling, or some other mechanism.  To our knowledge,
no other explanation has been advanced for these anamolous degeneracies.

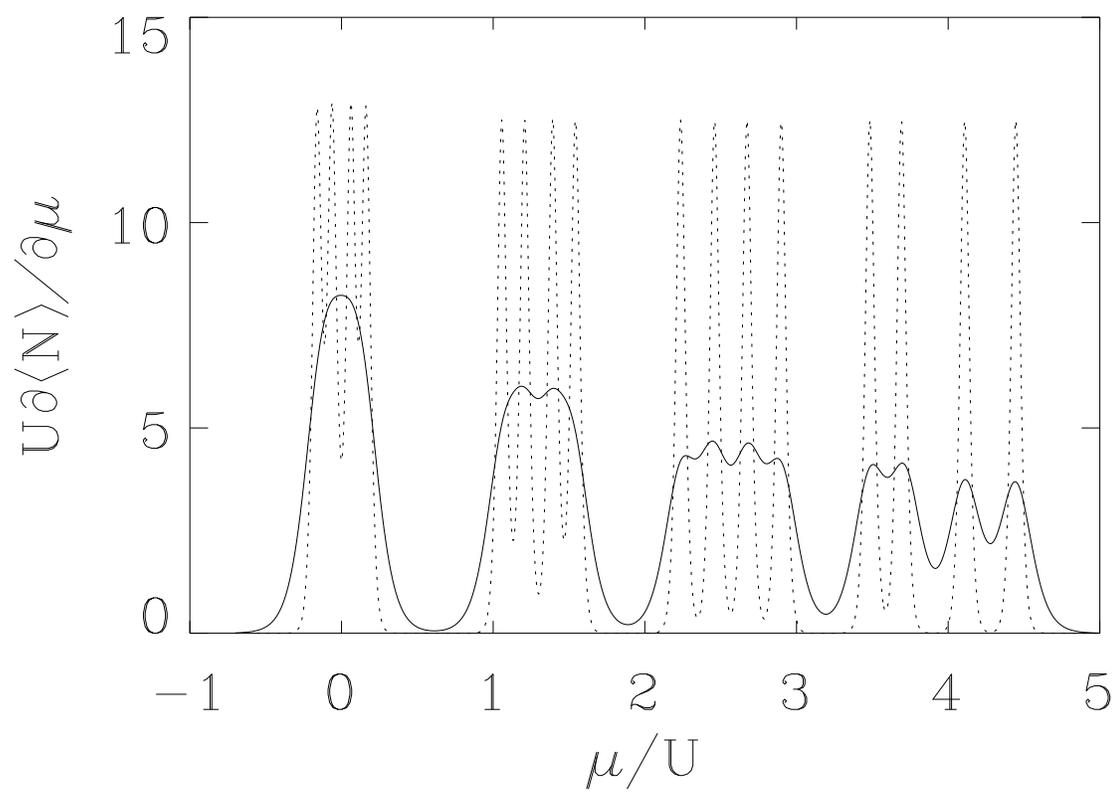
\begin{figure}
\caption{A plot of the electron addition spectrum
$\partial \langle \hat{N} \rangle / \partial \mu$
versus the chemical potential $\mu$ for a linear array of four quantum
dots with four single-particle energy levels per dot.  The level
spacing is $\Delta=0.3 U$, and the tunneling matrix elements are
$t_{\alpha}=0.1 U (1.5)^{\alpha -1}$, $\alpha=1,\ldots,4$.  Solid curve:
$kT=.07 U$.  Dotted curve: $kT=.02 U$.}
\label{fig3}
\end{figure}

Finally, we consider the effects of energy-dependent tunneling.  To this
point, we have assumed equal tunneling matrix elements for each of the
quantum dot states, but in fact the more weakly bound states should be
connected by stronger tunneling matrix elements.  To simulate this
effect, in Fig.~\ref{fig3} we show the electron addition spectrum for
a linear array of four quantum dots with four single-particle energy
levels per dot with tunneling matrix elements $t_{\alpha} = 0.1 U
(1.5)^{\alpha -1}$, $\alpha=1,\ldots,4$.  The level spacing is $\Delta =
0.3 U$. Note that for a linear array with open boundary
conditions, the degeneracies discussed above do not occur.
When disorder was
introduced into the tunneling matrix elements, the gross features of the
spectrum were unchanged, but the peak spacings within the
minibands were different.
The energy gap between the minibands decreases with
increasing chemical potential; when the value of $U$ was decreased 40\%
relative to $\Delta$ and $t_{\alpha}$, the energy gap between the third
and fourth minibands was no longer discernible.  As discussed above,
this breakdown of CCB \cite{spindegen}
is analogous to the Mott-Hubbard insulator-metal
transition \cite{mott,meanfield,andyandi}.
Quantum dot arrays should thus provide a
system where this transition can be studied by {\em tuning a gate voltage},
which would be inconceivable in the metal-oxide systems which display
conventional Mott-Hubbard metal-insulator transitions.

A reexamination of the work of Kouwenhoven {\it et al}.\ \cite{1dcrystal}
leads us to conclude that the analogue of the Mott-Hubbard
metal-insulator transition may have already been observed in a
1D array of quantum dots.  In Ref.~\cite{1dcrystal}, the conductance
spectrum of a linear array of fifteen GaAs quantum dots
$\sim 100\mbox{nm}$ in diameter was measured.
While our capacitance spectra cannot be compared
directly to those conductance spectra, the peak positions should be the
same in linear response, though the peak heights and possibly the
line shapes would be different.  At low values of the gate voltage
$V_{g2}$ (corresponding to low values of the chemical potential in the
array), conductance peaks separated by 4mV were seen in
Ref.~\cite{1dcrystal}.  Given the electrostatic ``lever arm'' of the gate
with respect to the quantum dots of about 4, this implies a 1mV spacing
in energy between the peaks, which is roughly the Coulomb blockade
energy for a 100nm GaAs quantum dot.  We therefore interpret these peaks
as Hubbard minibands in which the individual discrete states are
unresolved due to a finite electron temperature ({\it c.f.}, the lowest
miniband in Fig.~\ref{fig3}).  As the gate voltage $V_{g2}$ was increased,
the conductance peaks were seen to broaden \cite{1dcrystal},
and some structure in the
peaks became observable; finally the gap structure with a period of
4mV was destroyed, which we interpret as the breakdown of CCB.
The similarity to Fig.~\ref{fig3} is striking.
Further experimental work is clearly warranted to map out more fully
the ``phase diagram'' of quantum dot arrays.

Some shortcomings of the model considered here [Eq.~(\ref{hubham})]
should be pointed out.
The tight-binding approximation employed here to treat
interdot coupling is probably not adequate to treat the strong-tunneling
regime where the Coulomb blockade energy gap has broken down completely
(the plateau region in the conductance spectra of Ref.~\cite{1dcrystal});
the analysis given in Refs.~\cite{1dcrystal,noninteractingdots}
of that regime is probably more appropriate.
The Hubbard-type interaction term in Eq.~(\ref{hubham}) is also a rather
severe idealization of the electron-electron interactions within the
quantum dots.  More realistic treatments of the electron-electron
interactions within a quantum dot \cite{dotinfield}
are necessary \cite{ray2} to treat the regime of strong magnetic fields, where
strongly-correlated precursors of the bulk-incompressible
fractional quantum Hall states are
expected to occur.  However, in the regime of weak tunneling and weak
magnetic fields which we are
primarily concerned with, we believe the collective phenomena in
semiconductor quantum dot arrays are adequately described by our model.

In conclusion, we have calculated the electron addition spectrum, or
differential capacitance, of arrays of 4 to 6 coupled semiconductor
quantum dots in and near the Coulomb blockade regime using a generalized
Hubbard model.  We argue on general grounds that the interplay of
quantum confinement, interdot tunneling, and strong intradot Coulomb
interactions leads to three distinct zero temperature phases in
quantum dot arrays which have clear signatures in the electron
addition spectrum:
for weak tunneling, the quantum dot states are
split into minibands but the Coulomb blockade of individual dots is
maintained; for intermediate tunneling strengths, the Coulomb blockade
of individual dots is destroyed but there remains an energy gap of
collective origin analogous to the energy gap in a Mott insulator; and
for strong tunneling, the Coulomb blockade is destroyed altogether.
The weak-tunneling phase is characterized by degeneracies in the
electron addition spectrum which are lifted by disorder and
by magnetic fields.
The intermediate phase we refer to as ``collective Coulomb blockade.''
The breakdown of collective Coulomb blockade in the strong tunneling
regime is analogous to the Mott-Hubbard insulator-metal transition
\cite{mott,meanfield,andyandi}.
We have argued that the phenomenology of an early
experiment on transport through a 1D array of quantum dots \cite{1dcrystal}
can be understood in this framework.
We believe that quantum dot arrays are rather ideal physical
systems to directly verify the predictions of Hubbard-type
strongly-correlated electron interaction models, albeit at very
different energy scales ($\sim \,$ meV) compared with what these models
were originally intended for ($\sim \,$ eV).

This work was supported by the National Science Foundation (NSF) and the
United States Office of Naval Research (US-ONR).

\begin{references}
\bibitem{spindegen} The breakdown of CCB is qualitatively similar in
systems with spin degeneracy.
\end{references}
\end{document}